\documentclass[aps,prd,twocolumn,showpacs,preprintnumber,
               nofootinbib,superscriptaddress]{revtex4-2}

\usepackage{amsmath,amssymb,amsfonts}
\usepackage{xcolor}
\usepackage{booktabs}
\usepackage{colortbl}
\usepackage{array}
\usepackage{multirow}
\usepackage{makecell}
\usepackage{graphicx}
\usepackage{bm}
\usepackage[colorlinks=true,
            linkcolor=blue,
            citecolor=blue,
            urlcolor=blue]{hyperref}

\definecolor{palegreen}{RGB}{180,235,180}
\definecolor{palered}{RGB}{255,175,175}
\definecolor{paleorange}{RGB}{255,220,160}
\definecolor{paleblue}{RGB}{190,215,255}
\definecolor{paleyellow}{RGB}{255,250,190}

\newcommand{\Hpp}{H^{\pm\pm}}
\newcommand{\HppL}{H_L^{\pm\pm}}
\newcommand{\HppR}{H_R^{\pm\pm}}
\newcommand{\mHpp}{m(H^{\pm\pm})}

\begin{document}

\title{Revised exclusion limits on doubly charged Higgs bosons
from a reanalysis of the ATLAS multi-lepton search
at $\sqrt{s} = 13$\,TeV}

\author{Kirtiman Ghosh}
\email{kirti.gh@gmail.com}
\affiliation{Institute of Physics, Bhubaneswar,
  Sachivalaya Marg, Sainik School Post,
  Bhubaneswar 751005, India}
\affiliation{Homi Bhabha National Institute,
  Training School Complex, Anushakti Nagar,
  Mumbai 400094, India}

\author{Arun Kumar Nayak}
\email{nayak@iopb.res.in}
\affiliation{Institute of Physics, Bhubaneswar,
  Sachivalaya Marg, Sainik School Post,
  Bhubaneswar 751005, India}
\affiliation{Homi Bhabha National Institute,
  Training School Complex, Anushakti Nagar,
  Mumbai 400094, India}

\author{Debabrata Sahoo}
\email{debabrata.s@iopb.res.in}
\affiliation{Institute of Physics, Bhubaneswar,
  Sachivalaya Marg, Sainik School Post,
  Bhubaneswar 751005, India}
\affiliation{Homi Bhabha National Institute,
  Training School Complex, Anushakti Nagar,
  Mumbai 400094, India}

\date{\today}
\begin{abstract}
The ATLAS search for pair-produced doubly charged Higgs bosons in
multi-lepton final states using the full Run~2 dataset
[Eur.\ Phys.\ J.\ C \textbf{83} (2023) 605] reports the strongest
limits to date on the mass of doubly charged scalars, driven by an essentially background-free four-lepton channel. We show that the
four-lepton signal efficiency implied by the auxiliary cutflow of
that analysis exceeds a strict, mass-independent upper bound
derived from the equal-branching-ratio assumption of the search,
the leptonic $\tau$ branching fractions, and the ATLAS lepton
reconstruction efficiencies. We show that this excess cannot be
explained by hadronic $\tau$ or jet misidentification without
invoking fake rates far above realistic values. We regenerate the
signal independently and recompute the exclusion limit, using the
corrected signal yields, the ATLAS background predictions and
uncertainties, and the same $CL_s$ procedure implemented in
\textsc{pyhf}. The resulting expected limit lies systematically
above the ATLAS expected limit, by roughly a factor of two or more. This
shifts the expected lower mass bound from $1065$~GeV to
$\sim950$~GeV in the left-right symmetric type-II seesaw model,
and from $880$~GeV to $\sim770$~GeV in the Zee--Babu model.
\end{abstract}


\maketitle

\section{Introduction}
\label{sec:intro}

The existence of doubly charged scalar bosons is a robust
prediction of a broad class of extensions of the Standard Model
(SM). They arise naturally in the type-II seesaw
mechanism~\cite{Magg:1980ut,Schechter:1980gr,Cheng:1980qt,Lazarides:1980nt,Mohapatra:1980yp}, where the scalar sector of
the SM is augmented by an $SU(2)_L$ triplet with hypercharge
$Y=2$, whose neutral component acquires a vacuum expectation
value and generates Majorana masses for the light neutrinos.
The doubly charged member of the triplet, $\Hpp$, couples
directly to same-sign pairs of charged leptons, with couplings
fixed by the neutrino mass matrix.
Doubly charged scalars are also present in left-right symmetric
models (LRSMs)~\cite{Pati:1974yy,Mohapatra:1974hk,Senjanovic:1975rk}. In the LRSM, the gauge group
$SU(2)_L\times SU(2)_R\times U(1)_{B-L}$ is broken by triplet
scalars $\Delta_L$ and $\Delta_R$, each of which contains a
doubly charged component, $\HppL$ and $\HppR$.
Further examples include the Zee--Babu radiative neutrino mass
model~\cite{Zee:1985id,Babu:1988ki}, in which a doubly charged
$SU(2)_L$ singlet $k^{\pm\pm}$ appears, the Georgi--Machacek
model~\cite{Georgi:1985nv,Chanowitz:1985ug}, and 3-3-1
models~\cite{Singer:1980sw,Pisano:1991ee}.

The ATLAS analysis whose results we reexamine in this work is
interpreted within two of these frameworks: the left-right
symmetric type-II seesaw model, in which the relevant state is
the $SU(2)_L$-triplet $\HppL$ (common to the canonical type-II
seesaw and the LRSM), and the Zee--Babu model, whose $k^{\pm\pm}$
shares the quantum numbers of the LRSM $\HppR$.
For sufficiently small triplet vacuum expectation value,
$v_\Delta$, the leptonic decays
$\Hpp\to\ell^\pm\ell^{\prime\pm}$ dominate over the
$\Hpp\to W^\pm W^\pm$ mode, and the doubly charged scalar
manifests as a striking same-sign dilepton resonance.

At hadron colliders, doubly charged scalars are pair produced
through the Drell--Yan process
$pp\to\gamma^*/Z^*\to H^{++}H^{--}$, and, when a singly charged
partner is present, in associated production
$pp\to W^{*\pm}\to H^{\pm\pm}H^{\mp}$.
The pair-production cross section is purely electroweak and
depends only on the $\Hpp$ mass and its gauge quantum numbers,
which makes the resulting bounds largely model independent within
the leptonic decay scenario.
Searches for same-sign lepton pairs from doubly charged scalar
decays have a long history, extending from LEP and the Tevatron
to dedicated analyses at the LHC.
Using $4.7$~fb$^{-1}$ of $\sqrt{s}=7$~TeV data, ATLAS excluded
$\HppL$ masses below $409$, $375$, and $398$~GeV in the
$e^\pm e^\pm$, $e^\pm\mu^\pm$, and $\mu^\pm\mu^\pm$ channels,
respectively, assuming a $100\%$ branching ratio into the
corresponding final state~\cite{ATLAS:2012hi}; the CMS
Collaboration performed a comparable search at the same
energy~\cite{CMS:2012dun}.
With $36.1$~fb$^{-1}$ of $\sqrt{s}=13$~TeV data, ATLAS extended
the exclusion for $\HppL$ up to $870$~GeV and for $\HppR$ up to
$760$~GeV, assuming a $100\%$ branching ratio into light-lepton
final states~\cite{ATLAS:2017xqs}. A complementary ATLAS search
targeting the $\Hpp\to W^\pm W^\pm$ decay channel, relevant at
larger $v_\Delta$, excluded masses up to
$350$~GeV~\cite{ATLAS:2021jol}.
The doubly charged scalar sector has likewise been the subject of
extensive phenomenological study, including production through
vector-boson and photon
fusion~\cite{Dutta:2014dba,Babu:2016rcr}, exclusive
processes~\cite{Duarte:2022xpm}, the low-mass and
compressed-spectrum regions of parameter
space~\cite{Ashanujjaman:2022ofg,Ashanujjaman:2023tlj}, and
prospects at the HL-LHC and future hadron and lepton
colliders~\cite{BhupalDev:2018tox,Ashanujjaman:2021txz,
Ashanujjaman:2022tdn}.

The most sensitive analysis in this class is the ATLAS search of
Ref.~\cite{ATLAS:2022pbd}, which uses the full Run~2 dataset of
$139$~fb$^{-1}$ at $\sqrt{s}=13$~TeV.
That analysis searches for pair-produced $\Hpp$ bosons decaying
to same-sign lepton pairs $\Hpp\to\ell^\pm\ell^{\prime\pm}$ with
$\ell,\ell^\prime=e,\mu,\tau$, reconstructing final states with
two, three, or four leptons, but retaining only electrons and
muons as reconstructed objects; leptonic $\tau$ decays therefore
enter the light-lepton final states while hadronic $\tau$ decays
do not.
Under the benchmark assumption that the six leptonic branching
ratios are equal,
\begin{equation}
  \mathcal{B}(\Hpp\to e^\pm e^\pm)
  = \mathcal{B}(\Hpp\to e^\pm \mu^\pm)
  = \cdots
  = \frac{1}{6},
  \label{eq:brassumption}
\end{equation}
the analysis reports an observed (expected) lower limit on the
$\HppL$ mass of $1080$~GeV ($1065^{+30}_{-50}$~GeV) within the
left-right symmetric type-II seesaw model, and $900$~GeV
($880^{+30}_{-40}$~GeV) within the Zee--Babu model.
The four-lepton channel, which is essentially background free,
drives the combined sensitivity and is responsible for the
strength of the quoted limits.

In the course of attempting to reproduce these results, we
identified an inconsistency in the signal efficiencies underlying
the four-lepton channel.
The auxiliary material~\cite{ATLAS:2022pbd_aux} of Ref.~\cite{ATLAS:2022pbd} provides a
detailed signal cutflow for the four-lepton final state.
The fraction of pair-production events retained at the
``four-lepton'' stages of this cutflow substantially exceeds the
maximum fraction that can be obtained analytically from the
branching-ratio assumption of Eq.~\eqref{eq:brassumption}
together with the known leptonic $\tau$ branching fractions and
the lepton reconstruction efficiencies quoted in the same paper.
Because the four-lepton channel drives the combined mass limit,
any overestimate of the four-lepton signal efficiency propagates
directly into an overly stringent bound on the production cross
section and hence on the $\Hpp$ mass.\footnote{No prior work has
identified this specific inconsistency, but two independent
reinterpretations of Ref.~\cite{ATLAS:2022pbd} show symptoms of the
same difficulty: Ref.~\cite{Bolton:2024thn} declines to recast its
combined four-lepton signal region, citing the complexity of the
equal-branching-ratio combination, while Ref.~\cite{Englert:2026eou}
calibrates its own recast to the ATLAS limit via an ad hoc
normalisation rather than reproducing it directly. This is
consistent with the four-lepton channel not being straightforwardly
reproducible from public information.}

In this work we quantify this inconsistency and assess its impact
on the exclusion limits. We first derive, analytically, the
maximum fraction of $pp\to H^{++}H^{--}$ events that can yield
four reconstructed light leptons under the equal-branching-ratio
assumption, accounting fully for leptonic $\tau$ decays and for
the lepton reconstruction and identification efficiencies of the
ATLAS analysis. We show that this fraction is a strict upper
bound lying well below the values implied by the auxiliary
cutflow, and that the excess cannot be accounted for by the
misidentification of hadronic $\tau$ decays or accompanying jets
as electrons without invoking unrealistic fake rates. We then
generate the signal independently using
\textsc{MadGraph5\_aMC@NLO} interfaced with a fast detector
simulation, implementing the object reconstruction and event
selection prescribed by ATLAS, and recompute the four-lepton
signal yields. Using these corrected yields together with the SM
background predictions reported by ATLAS, we re-derive the upper
limit on the $\Hpp$ pair-production cross section following the
same statistical procedure, implemented in \textsc{pyhf}. The
resulting exclusion is significantly weaker than that quoted in
Ref.~\cite{ATLAS:2022pbd}.

The rest of the paper proceeds as follows. Section~\ref{sec:analytical}
derives a model-independent upper bound on the four-lepton signal
efficiency and shows that it conflicts with the cutflow numbers in
the ATLAS auxiliary material, along with an assessment of whether
fake leptons could account for the discrepancy. We then turn, in
Sec.~\ref{sec:simulation}, to our own implementation of the signal
generation, reconstruction, and event selection, benchmarking it
directly against the ATLAS auxiliary cutflow at each stage. The
re-derived cross-section limits are presented in
Sec.~\ref{sec:results}, together with a lepton-multiplicity-by-
multiplicity comparison to the ATLAS signal yields and the
normalisation hint that emerges from it. Section~\ref{sec:conclusion}
closes with our conclusions.

\section{Theoretical constraint on the four-lepton signal efficiency}
\label{sec:analytical}

In this section we compute the fraction of $pp\to H^{++}H^{--}$
pair-production events that can produce four reconstructed light
leptons ($e$ or $\mu$) in the final state, under the
equal-branching-ratio assumption of Eq.~\eqref{eq:brassumption}.
The calculation rests only on the stated branching ratios, the $\tau$ leptonic branching fractions, and the lepton
reconstruction efficiencies quoted by ATLAS, and yields a strict
upper bound on the four-lepton signal efficiency that is
independent of the $\Hpp$ mass.

\subsection{Effective per-arm two-light-lepton probabilities}
\label{sec:perarm}

We treat each $\Hpp$ as an independent decay ''arm''. A light
lepton in the final state originates either directly, from a
decay mode $\Hpp\to e^\pm e^\pm, e^\pm\mu^\pm$, or $\mu^\pm\mu^\pm$,
or indirectly, from a leptonic $\tau$ decay in one of the three
modes containing a $\tau$ ($\Hpp\to e^\pm\tau^\pm,
\mu^\pm\tau^\pm, \tau^\pm\tau^\pm$). The relevant leptonic $\tau$
branching fractions are~\cite{ParticleDataGroup:2022pth}
\begin{align}
  b_e   &\equiv \mathcal{B}(\tau\to e\,\nu_e\bar\nu_\tau)
         = 0.18, \\
  b_\mu &\equiv \mathcal{B}(\tau\to\mu\,\nu_\mu\bar\nu_\tau)
         = 0.17, \\
  b_\ell &\equiv b_e + b_\mu = 0.35 .
\end{align}
For a single arm we define $f(\ell_1\ell_2)$ as the probability
of producing a same-sign light-lepton pair
$(\ell_1,\ell_2)$ with $\ell_1,\ell_2\in\{e,\mu\}$, summed over
all six decay modes and including subsequent leptonic $\tau$
decays. Summing the contributions from each of the six
equiprobable decay modes, including those where one or both
leptons arise from a subsequent leptonic $\tau$ decay, gives
\begin{align}
  f(ee)   &= \tfrac{1}{6}\left(1 + b_e + b_e^2\right) = 0.20, \\
  f(e\mu) &= \tfrac{1}{6}\left(1 + b_e + b_\mu + 2b_eb_\mu\right)
             = 0.24, \\
  f(\mu\mu) &= \tfrac{1}{6}\left(1 + b_\mu + b_\mu^2\right)
             = 0.20 .
\end{align}
The total probability that an arm yields two light leptons is
\begin{equation}
  \Sigma f_2 \equiv f(ee)+f(e\mu)+f(\mu\mu)
  = \tfrac{1}{6}\!\left(3 + 2b_\ell + b_\ell^2\right)
  = 0.64 .
  \label{eq:sigmaf2}
\end{equation}

\subsection{Truth-level four-light-lepton fraction}

The four-light-lepton fraction is given by the square of the
per-arm probability,
\begin{equation}
  P_4^{\rm truth} = \left(\Sigma f_2\right)^2
  = (0.64)^2 = 0.41 .
  \label{eq:p4truth}
\end{equation}
This bound follows solely from Eq.~\eqref{eq:brassumption} and
the leptonic $\tau$ branching fractions; it is independent of
$\mHpp$ and precedes any detector modelling. The channel-by-channel
decomposition in Table~\ref{tab:truth_reco} is essential for the
next step: since $\epsilon_e \neq \epsilon_\mu$, the
reconstruction efficiency in Sec.~\ref{sec:reco} must be applied
separately to each of the five $(n_e,n_\mu)$ channels rather than
to the total.

\subsection{Fraction after reconstruction efficiency}
\label{sec:reco}

The ATLAS analysis quotes reconstruction and identification
efficiencies for the tight lepton selection of
$\epsilon_e = 0.88$ for electrons and $\epsilon_\mu = 0.95$ for
muons in the relevant kinematic range~\cite{ATLAS:2022pbd}. Applying
these per-lepton efficiencies to each channel gives the fraction
of events with four reconstructed light leptons,
\begin{equation}
  P_4^{\rm reco} = \sum_{n_e+n_\mu=4}
     P_4^{\rm truth}(n_e,n_\mu)\,
     \epsilon_e^{\,n_e}\epsilon_\mu^{\,n_\mu} = 0.29 ,
  \label{eq:p4reco}
\end{equation}
as detailed in Table~\ref{tab:truth_reco}. Equivalently, defining
the per-arm reconstruction efficiency
\begin{equation}
  \bar\epsilon_{\rm arm}
  = \frac{f(ee)\,\epsilon_e^2 + f(e\mu)\,\epsilon_e\epsilon_\mu
         + f(\mu\mu)\,\epsilon_\mu^2}{\Sigma f_2}
  = 0.84 ,
  \label{eq:earm}
\end{equation}
one has $P_4^{\rm reco} = (\Sigma f_2)^2\,\bar\epsilon_{\rm arm}^2
= P_4^{\rm truth}\,\bar\epsilon_{\rm arm}^2$.

\begin{table}[t]
  \centering
  \caption{Truth-level and post-reconstruction four-light-lepton
  fractions by lepton-flavour channel, including leptonic
  $\tau$ decays and the ATLAS lepton efficiencies
  ($\epsilon_e=0.88$, $\epsilon_\mu=0.95$), prior to any
  kinematic selection.}
  \label{tab:truth_reco}
  \renewcommand{\arraystretch}{1.3}
  \begin{tabular}{lccc}
    \toprule
    Channel & Truth & Reco.\ eff. & Reco. \\
    \midrule
    $4e$        & $0.04$  & $0.60$ & $0.02$ \\
    $3e\,1\mu$  & $0.10$  & $0.65$ & $0.06$ \\
    $2e\,2\mu$  & $0.14$  & $0.70$ & $0.10$ \\
    $1e\,3\mu$  & $0.09$  & $0.75$ & $0.07$ \\
    $4\mu$      & $0.04$  & $0.81$ & $0.03$ \\
    \midrule
    {Total} & $0.41$ & $0.70$ & ${0.29}$ \\
    \bottomrule
  \end{tabular}
\end{table}

The value $P_4^{\rm reco} = 0.29$ is itself an upper bound on the
observable four-lepton fraction, since it precedes the kinematic
acceptance requirements of the analysis.
This gives the strict hierarchy
\begin{equation}
  P_4^{\rm obs} \;\leq\; P_4^{\rm reco} = 0.29
  \;\leq\; P_4^{\rm truth} = 0.41 ,
  \label{eq:hierarchy}
\end{equation}
with the inequalities expected to be strict in practice.

\subsection{Comparison with the ATLAS auxiliary cutflow}
\label{sec:comparison}

Auxiliary Table~5 \cite{ATLAS:2022pbd_aux} of Ref.~\cite{ATLAS:2022pbd} lists the cumulative
four-lepton signal yields at each selection stage, normalised to
$139$~fb$^{-1}$, for signal masses of $700$, $900$, $1100$, and
$1300$~GeV. Dividing the ``four loose leptons'' and ``four tight
leptons'' entries by the total ``Yield'' entry gives retained
fractions of $0.50$--$0.51$ (loose) and $0.43$--$0.46$ (tight)
across the four benchmark masses -- both above the truth-level
maximum $P_4^{\rm truth}=0.41$ and far above the
reconstruction-level bound $P_4^{\rm reco}=0.29$. One feature is
noteworthy: the ``four tight'' fraction exceeds even the
truth-level bound, so the discrepancy cannot be attributed to
imperfect modelling of lepton efficiencies, which can only lower
the fraction.

\subsection{Can hadronic $\tau$ or jet misidentification
account for the excess?}
\label{sec:fake}

A natural candidate for populating the four-lepton sample beyond
the genuine four-light-lepton events is the class of events with
exactly three genuine light leptons, in which a hadronic $\tau$
or an accompanying jet is misidentified as an electron. We now
show that the fake rate required to bridge the gap is far larger
than any realistic value.

Events with exactly three genuine light leptons at truth level
arise when one arm produces two light leptons and the other
produces a single light lepton, the remaining $\tau$ decaying
hadronically. The per-arm probability of yielding exactly one
light lepton is
\begin{align}
  f(e)   &= \tfrac{1}{6}(1-b_\ell)(1+2b_e) = 0.15, \\
  f(\mu) &= \tfrac{1}{6}(1-b_\ell)(1+2b_\mu) = 0.15, \\
  \Sigma f_1 &= f(e)+f(\mu) = 0.29 ,
\end{align}
so that the truth-level three-light-lepton fraction is
\begin{equation}
  P_3^{\rm truth} = 2\,\Sigma f_2\,\Sigma f_1 = 0.37 .
  \label{eq:p3truth}
\end{equation}
Crucially, every such event contains exactly one hadronic $\tau$,
originating from the single-lepton arm.

For a three-lepton event to enter the apparent four-lepton
sample, the three genuine leptons must be reconstructed and the
hadronic $\tau$ (or an accompanying jet) must fake an electron.
The reconstruction efficiency for the three genuine leptons is
$\epsilon_{3\rm gen} = \bar\epsilon_{\rm arm}\,\bar\epsilon_1$,
where $\bar\epsilon_{\rm arm}=0.84$ from Eq.~\eqref{eq:earm} and
\begin{equation}
  \bar\epsilon_1
  = \frac{f(e)\,\epsilon_e + f(\mu)\,\epsilon_\mu}{\Sigma f_1}
  = 0.91 ,
\end{equation}
giving $\epsilon_{3\rm gen} = 0.77$. The contribution of these
events to the apparent four-lepton sample, per produced pair, is
\begin{equation}
  \frac{N_{4\ell}^{\rm fake}}{N_{\rm prod}}
  = P_3^{\rm truth}\,\epsilon_{3\rm gen}\,\epsilon_{\tau_h\to e}
  = 0.29\,\epsilon_{\tau_h\to e} ,
  \label{eq:fakefrac}
\end{equation}
where $\epsilon_{\tau_h\to e}$ is the probability that the
hadronic $\tau$ is reconstructed as a tight electron. Requiring
the total apparent four-tight-lepton fraction to match the ATLAS
value of $\approx0.45$,
\begin{equation}
  0.29 + 0.29\,\epsilon_{\tau_h\to e} = 0.45 ,
\end{equation}
yields
\begin{equation}
  \epsilon_{\tau_h\to e}^{\rm tight} = 0.55 .
  \label{eq:tightfake}
\end{equation}

The required mis-tagging rate is far above realistic values: ATLAS
electron identification rejects hadronic $\tau$ decays and jets
at the percent level or below~\cite{ATLAS:2019qmc,Flechl:2017bse},
roughly an order of magnitude smaller than the $55\%$ needed here. We stress that these are conservative lower bounds: the
calculation leading to Eq.~\eqref{eq:tightfake} uses $P_4^{\rm
reco}$ and $\epsilon_{3\rm gen}$, both evaluated before the
kinematic acceptance cuts ($p_T$, $\eta$, and isolation) of the
ATLAS selection are applied. Since these cuts can only reduce the
genuine three- and four-lepton yields entering the estimate, the
true fake rate needed to explain the ATLAS excess is larger still
than the $55\%$ quoted above.

Including accompanying ISR jets enlarges the pool of objects that
could fake an electron, but since the four-lepton region requires
exactly four reconstructed leptons, only events with exactly one
fake among the $N_j=1+n_{\rm ISR}$ candidates contribute; the
relevant probability is
$N_j\,\epsilon_{j\to e}(1-\epsilon_{j\to e})^{N_j-1}$. As a
function of $\epsilon_{j\to e}$, this quantity is maximised at
$\epsilon_{j\to e}=1/N_j$, where it attains the value
$(1-1/N_j)^{N_j-1}$, which falls from $1$ at $N_j=1$ toward
$1/e\approx0.37$ as $N_j$ grows. For the tight selection, the
required contamination ($\approx0.55$) exceeds this maximum
achievable value for any $N_j\geq2$: no per-jet fake rate can
reproduce the excess once more than one fake-capable object is
present. 

We conclude that neither hadronic $\tau$ misidentification nor
jet misidentification, alone or in combination, can account for
the excess of the ATLAS four-lepton fractions over the
theoretical bound of Eq.~\eqref{eq:hierarchy}.

\section{Signal simulation and event selection}
\label{sec:simulation}

To assess the impact of the four-lepton signal efficiency on the
exclusion limit independently of the auxiliary cutflow, we
generated the signal from first principles and applied a detector
simulation and event selection reproducing, as closely as
possible, the procedure of Ref.~\cite{ATLAS:2022pbd}.

Pair production of doubly charged scalars,
$pp\to H^{++}H^{--}$, was simulated at leading order with
\textsc{MadGraph5\_aMC@NLO}~\cite{Alwall:2014hca} using the
type-II seesaw model implementation, with the
$H_L^{\pm\pm}$ couplings to leptons chosen to realise the
equal-branching-ratio benchmark of
Eq.~\eqref{eq:brassumption}. Signal samples were generated for
$\Hpp$ masses in the range $400$--$1300$~GeV. The NLO-to-LO
$K$-factors and the reference cross sections were taken from
Refs.~\cite{Fuks:2019clu,Ruiz:2022sct} so that our production
normalisation matches that used by ATLAS \cite{ATLAS:2022pbd_aux}. Parton showering and
hadronisation were performed with \textsc{Pythia\,8}%
~\cite{Sjostrand:2014zea}, and the detector response was modelled
with \textsc{Delphes\,3}~\cite{deFavereau:2013fsa} using a card
tuned to reproduce the ATLAS lepton reconstruction and
identification efficiencies, isolation performance, and $\eta$
acceptance.

Object reconstruction and event selection follow Sec.~4 of
Ref.~\cite{ATLAS:2022pbd}. Electron candidates are required to have
$p_T > 40$~GeV and $|\eta| < 2.47$, excluding the calorimeter
transition region $1.37 < |\eta| < 1.52$, and to satisfy tight
identification and isolation. Muon candidates are required to
have $p_T > 40$~GeV and $|\eta| < 2.5$. Jets are reconstructed with the anti-$k_t$
algorithm~\cite{Cacciari:2008gp} with radius parameter $R=0.4$
and are required to have $p_T > 20$~GeV and $|\eta|<2.5$; events
containing $b$-tagged jets are vetoed. In our simulation,
electron and muon reconstruction and identification are modelled
using the $p_T$- and $\eta$-dependent efficiency maps from the
ATLAS electron~\cite{ATLAS:2019qmc,ATLAS:2019jvq} and
muon~\cite{ATLAS:2020auj,ATLAS:2016lqx} performance measurements, the
same references used by Ref.~\cite{ATLAS:2022pbd} itself.

The four-lepton signal region is defined following
Ref.~\cite{ATLAS:2022pbd}: events are required to contain two
same-charge light-lepton pairs with zero net charge, with a
$b$-jet veto and the $Z$-veto applied, and with the average
invariant mass of the two same-charge pairs,
$\overline{M} = (m_{\ell^+\ell^{\prime+}}
+ m_{\ell^-\ell^{\prime-}})/2$, required to exceed $300$~GeV.
The leading same-charge pair is further required to satisfy
$m(\ell^\pm,\ell^{\prime\pm})_{\rm lead} > 300$~GeV.
\begin{table}[t]
  \centering
  \caption{Four-lepton signal-region cutflow: ATLAS auxiliary
  Table 5 \cite{ATLAS:2022pbd_aux} of Ref.~\cite{ATLAS:2022pbd} compared, stage by stage, with
  the corresponding yields from our own signal simulation, for
  four $\Hpp$ mass points. The ratio column is Ours/ATLAS.}
  \label{tab:cutflow_compare}
  \renewcommand{\arraystretch}{1.15}
  \begin{tabular}{llccc}
    \toprule
    $\mHpp$ & Stage & ATLAS & Ours & Ratio \\
    \midrule
    \multirow{7}{*}{700 GeV}
      & Yield                     & 68   & 68   & 1.00 \\
      & Four tight leptons        & 31   & 11.4 & 0.36 \\
      & Correct charge comb.      & 29   & 10.7 & 0.36 \\
      & $b$-jet veto              & 28   & 10.5 & 0.37 \\
      & $m(\ell^\pm,\ell'^\pm)_{\rm lead}$ & 28 & 10.4& 0.37 \\
      & $\overline{M}$            & 28   & 10.4 & 0.37 \\
      & $Z$-veto                  & 28   & 10.1 & 0.36 \\
    \midrule
    \multirow{7}{*}{900 GeV}
      & Yield                     & 16   & 16  & 1.00 \\
      & Four tight leptons        & 7.2  & 2.6 & 0.36 \\
      & Correct charge comb.      & 6.6  & 2.5 & 0.37 \\
      & $b$-jet veto              & 6.4  & 2.4 & 0.37 \\
      & $m(\ell^\pm,\ell'^\pm)_{\rm lead}$ & 6.4 & 2.4 & 0.37 \\
      & $\overline{M}$            & 6.4  & 2.4 & 0.37 \\
      & $Z$-veto                  & 6.3  & 2.4 & 0.38 \\
    \midrule
    \multirow{7}{*}{1100 GeV}
      & Yield                     & 4.2  & 4.2 & 1.00 \\
      & Four tight leptons        & 1.9  & 0.72 & 0.38 \\
      & Correct charge comb.      & 1.7  & 0.67 & 0.39 \\
      & $b$-jet veto              & 1.7  & 0.66 & 0.39 \\
      & $m(\ell^\pm,\ell'^\pm)_{\rm lead}$ & 1.7 & 0.66 & 0.39 \\
      & $\overline{M}$            & 1.7  & 0.66 & 0.39 \\
      & $Z$-veto                  & 1.6  & 0.65 & 0.40 \\
    \midrule
    \multirow{7}{*}{1300 GeV}
      & Yield                     & 1.2  & 1.2 & 1.00 \\
      & Four tight leptons        & 0.52 & 0.20 & 0.38 \\
      & Correct charge comb.      & 0.47 & 0.18 & 0.38 \\
      & $b$-jet veto              & 0.45 & 0.18 & 0.40 \\
      & $m(\ell^\pm,\ell'^\pm)_{\rm lead}$ & 0.45 & 0.18 & 0.40 \\
      & $\overline{M}$            & 0.45 & 0.18 & 0.40 \\
      & $Z$-veto                  & 0.45 & 0.18 & 0.40 \\
    \bottomrule
  \end{tabular}
\end{table}

Applying this selection to our signal samples, we obtain the
four-lepton cutflow given in Table~\ref{tab:cutflow_compare},
alongside the corresponding stage-by-stage ATLAS yields from the
auxiliary cutflow ~\cite{ATLAS:2022pbd_aux} of Ref.~\cite{ATLAS:2022pbd}. At every mass point,
our simulation yields a retained four-lepton fraction consistent
with the reconstruction-level bound $P_4^{\rm reco}=0.29$ derived
in Sec.~\ref{sec:analytical}, in contrast to the $0.38$--$0.41$
retained by ATLAS after the full selection.

It is important to note that in all four
benchmark masses, the ratio of our simulated event count to the
corresponding ATLAS entry is approximately constant, rising slowly from $\sim0.36$--$0.37$ at
$\mHpp=700$~GeV to $\sim0.38$--$0.40$ at $\mHpp=1300$~GeV\footnote{The mass dependence has a kinematic origin:
$P_4^{\rm truth}=0.41$ splits into direct decays ($0.25$) and
$\tau$-mediated decays ($0.16$), and leptons from the latter are
softer and less likely to pass the $p_T$ and
$m(\ell^\pm,\ell'^\pm)_{\rm lead}$ thresholds, an effect most
severe at low $\mHpp$ where the $\tau$ is least boosted. As
$\mHpp$ increases the $\tau$-decay leptons harden and the
retained fraction approaches $P_4^{\rm truth}$ from below,
matching the $0.36\to0.40$ trend in
Table~\ref{tab:cutflow_compare}.}. This
ratio is numerically close to, but systematically below,
$P_4^{\rm truth}=(\Sigma f_2)^2=0.41$ (see
Eq.~\eqref{eq:p4truth}). This near-constant Ours/ATLAS ratio,
sitting just below $P_4^{\rm truth}$ at every mass point, hints
towards a sample normalisation issue in the ATLAS four-lepton
signal sample. 

\begin{figure}[!t]
  \centering
  \includegraphics[width=0.8\columnwidth]{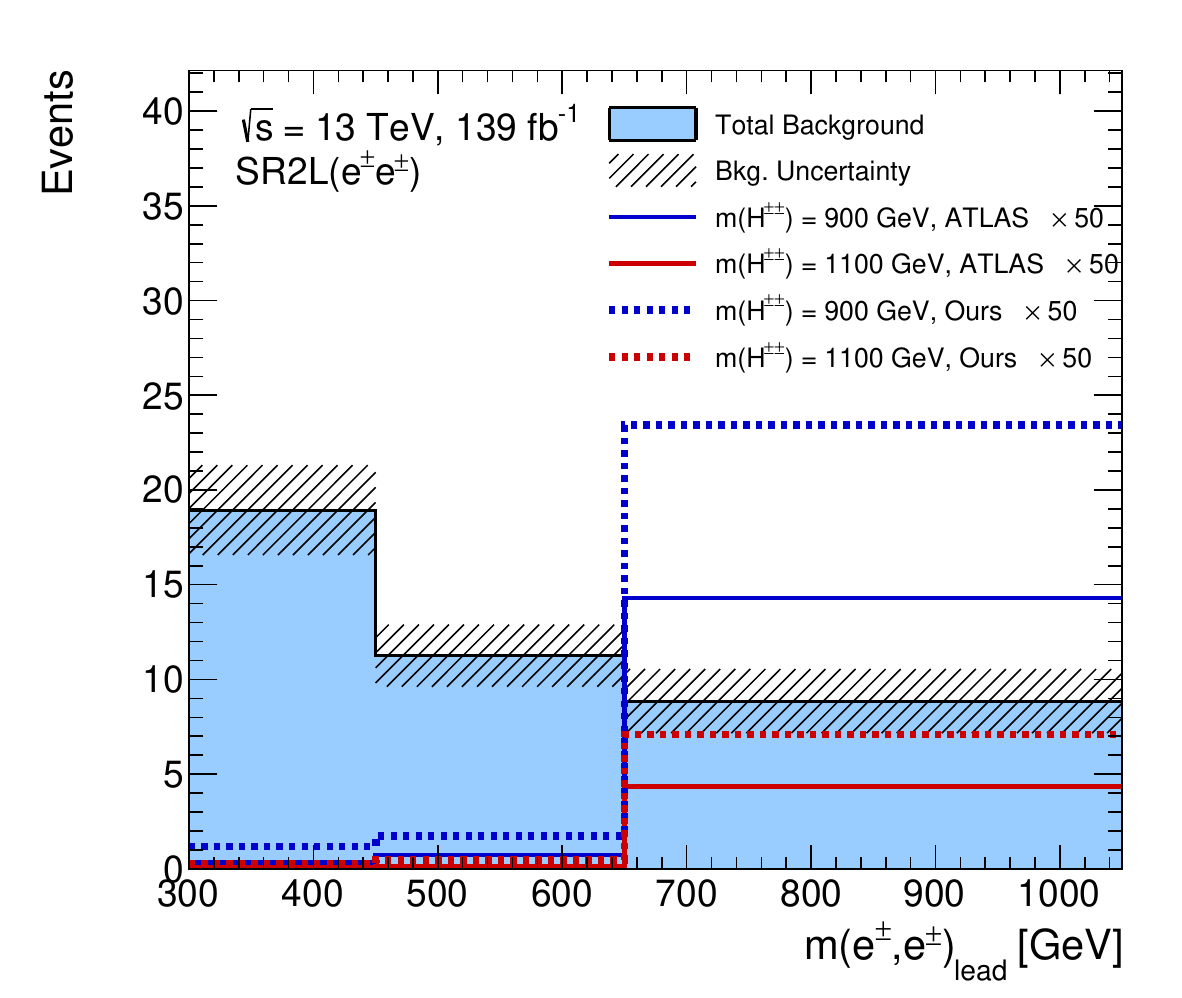}\\
  \includegraphics[width=0.8\columnwidth]{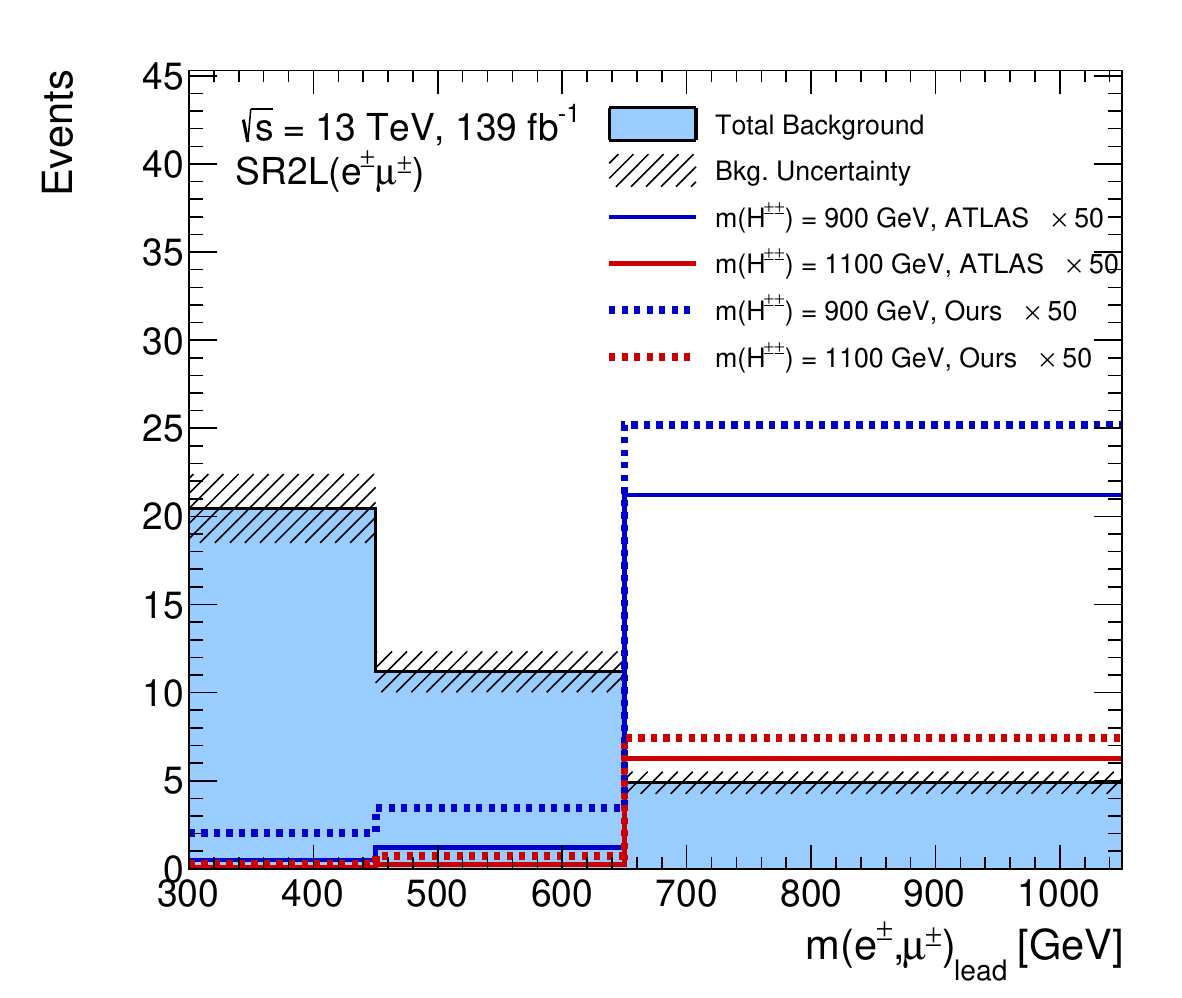}\\
  \includegraphics[width=0.8\columnwidth]{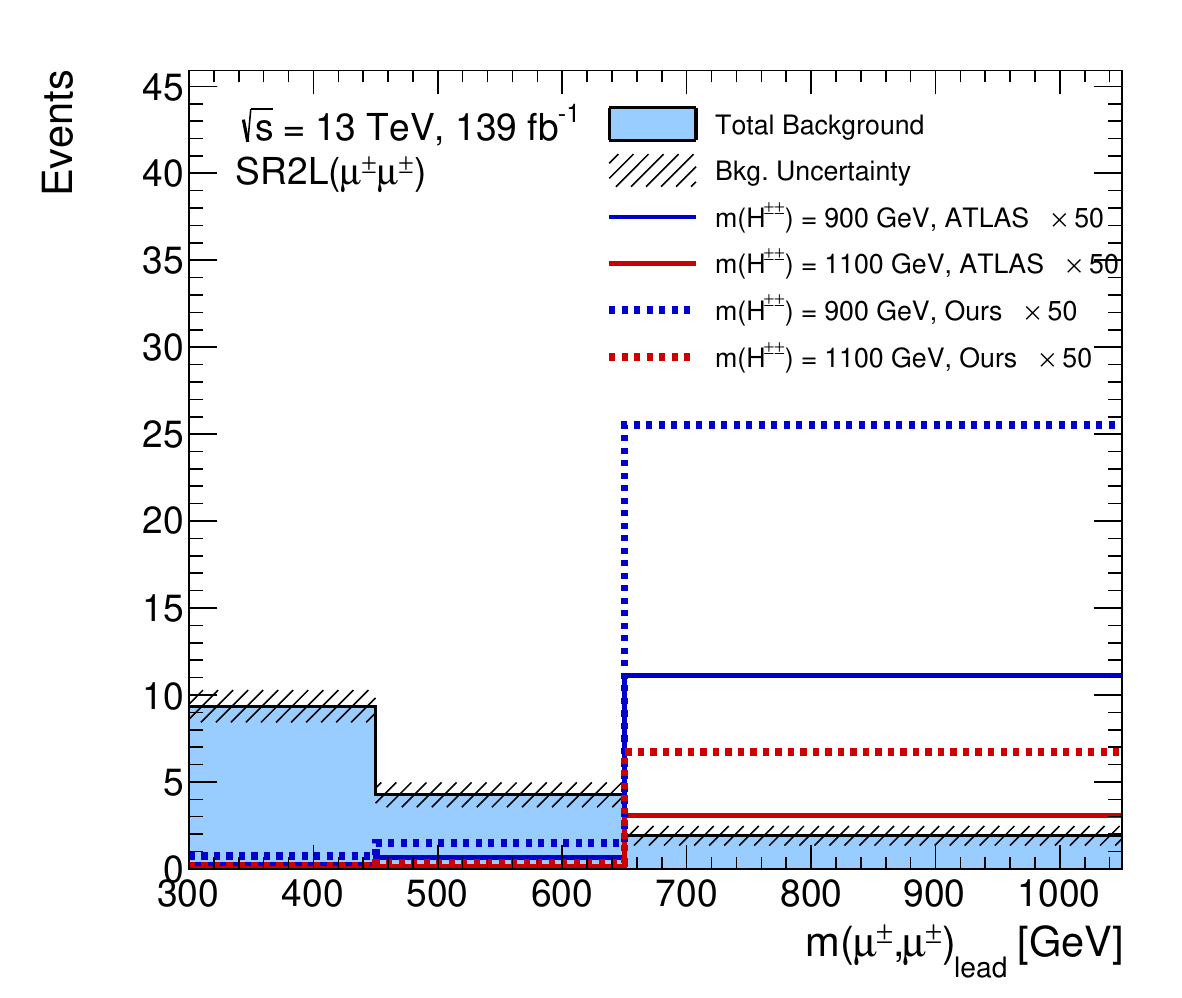}
  \caption{Distributions of the leading same-sign dilepton
  invariant mass, $m(\ell^\pm,\ell'^\pm)_{\rm lead}$, in the
  two-lepton signal region, for the (top panel) $e^\pm e^\pm$, (middle panel)
  $e^\pm\mu^\pm$, and (bottom panel) $\mu^\pm\mu^\pm$ channels. The stacked
  histogram shows the ATLAS post-fit SM background. The solid
  lines show the ATLAS signal prediction and the dotted lines
  show our own simulated signal.}
  \label{fig:sr2l_compare}
\end{figure}

\section{Revised cross-section limits}
\label{sec:results}

 Thus far we have focused
exclusively on the four-lepton final state, for two reasons.
First, this channel is almost entirely background free and
therefore drives the combined exclusion limit. Second, a
theoretical upper bound on its signal efficiency is comparatively
straightforward to derive, since it depends only on the assumed
branching ratios and the lepton identification efficiencies. The
two- and three-lepton final states, by contrast, receive
contributions from higher-multiplicity events in which one or
more leptons go unidentified or fall outside the detector
acceptance, so an analogous theoretical bound would additionally
require modelling the kinematic acceptance efficiency for each
such contribution.
For this reason, we used the four-lepton channel as a proof of
concept to establish the normalisation discrepancy identified in
this work. Whether a similar effect is present in the other
signal regions of Ref.~\cite{ATLAS:2022pbd} cannot be established with
the same rigour, given the added complexity noted above; we
nonetheless use our independently simulated signal in all signal
regions, on the same footing as in the four-lepton case, to
obtain the revised combined exclusion limit on the doubly charged
Higgs pair-production cross section as a function of $\mHpp$.

We recompute the exclusion limit using precisely the statistical
procedure of Ref.~\cite{ATLAS:2022pbd}. The ATLAS signal regions are
defined as follows: SR2L requires two same-charge
light leptons with $m(\ell^\pm,\ell'^\pm)_{\rm lead}>300$~GeV,
$\Delta R(\ell^\pm,\ell'^\pm)<3.5$, and
$p_T(\ell^\pm,\ell'^\pm)_{\rm lead}>300$~GeV; SR3L requires three
leptons with the same $m(\ell^\pm,\ell'^\pm)_{\rm lead}$ and
$p_T(\ell^\pm,\ell'^\pm)_{\rm lead}$ requirements; and SR4L
requires four leptons forming two same-charge, zero-net-charge
pairs, with $m(\ell^\pm,\ell'^\pm)_{\rm lead}>300$~GeV and
$\overline{M}>300$~GeV. A $b$-jet veto is applied in all regions,
and a $Z$-veto is additionally applied in the three- and
four-lepton regions.

The statistical analysis of Ref.~\cite{ATLAS:2022pbd} implements a
binned maximum-likelihood fit using HistFitter, with
$m(\ell^\pm,\ell'^\pm)_{\rm lead}$ as the fit variable in SR2L and
SR3L, and a single-bin event yield in SR4L. A $95\%$~CL upper limit
on $\sigma(pp\to H^{++}H^{--})$ is obtained using the $CL_s$
method~\cite{Read:2002hq}, with pseudo-experiments used in place
of the asymptotic approximation because of the small yields in
some regions.

\begin{figure}[!t]
  \centering
  \includegraphics[width=0.8\columnwidth]{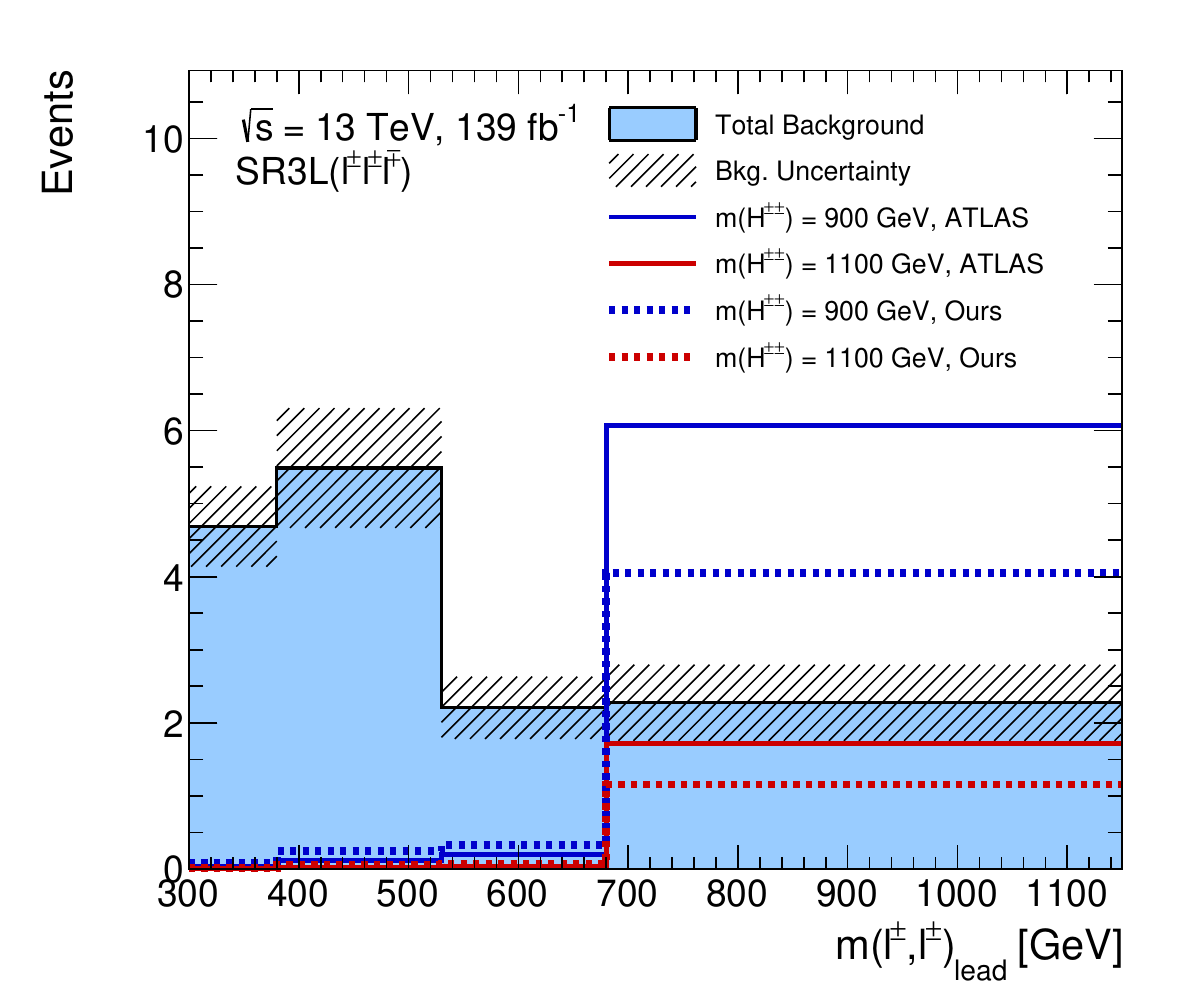}
  \caption{Distribution of the leading same-sign dilepton
  invariant mass, $m(\ell^\pm,\ell'^\pm)_{\rm lead}$, in the
  three-lepton signal region (SR3L). The stacked histogram shows
  the ATLAS post-fit SM background. The solid lines show the
  ATLAS signal prediction and the dotted lines show our own
  simulated signal, for $\mHpp=900$ and $1100$~GeV.}
  \label{fig:sr3l_compare}
\end{figure}

\begin{figure}[!t]
  \centering
  \includegraphics[width=0.8\columnwidth]{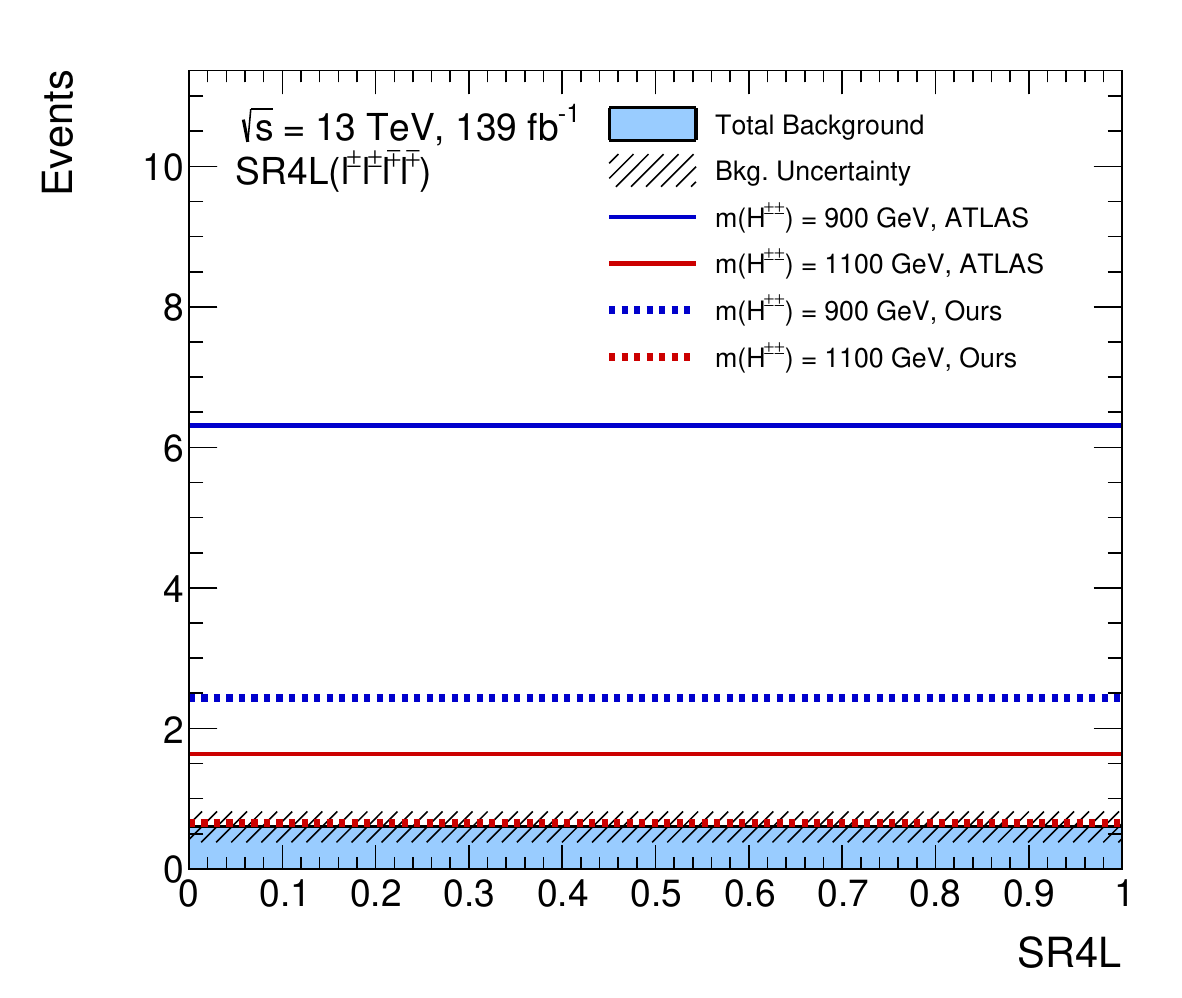}
  \caption{Event yield in the four-lepton signal region (SR4L). The
  stacked histogram shows the ATLAS post-fit SM background. The
  solid lines show the ATLAS signal prediction and the dotted
  lines show our own simulated signal, for $\mHpp=900$ and
  $1100$~GeV.}
  \label{fig:sr4l_compare}
\end{figure}

We reproduce this procedure exactly, using \textsc{pyhf}%
~\cite{Feickert:2022lzh,Heinrich:2021gyp} for the statistical fit. We adopt the
ATLAS post-fit background predictions and the uncertainties directly from
Ref.~\cite{ATLAS:2022pbd} rather than re-deriving them, since our
results concern only the signal side of the fit. Figure~\ref{fig:sr2l_compare}, Figure~\ref{fig:sr3l_compare}, and
Figure~\ref{fig:sr4l_compare} present, respectively, the
$m(\ell^\pm,\ell'^\pm)_{\rm lead}$ distributions in SR2L (all
three flavour channels), SR3L, and SR4L, with our simulated signal
overlaid on the ATLAS post-fit background and ATLAS signal prediction. Our simulated signal distributions are the ones used as input to
the \textsc{pyhf} likelihood fit.

\begin{figure}[t]
  \centering
  \includegraphics[width=\columnwidth]{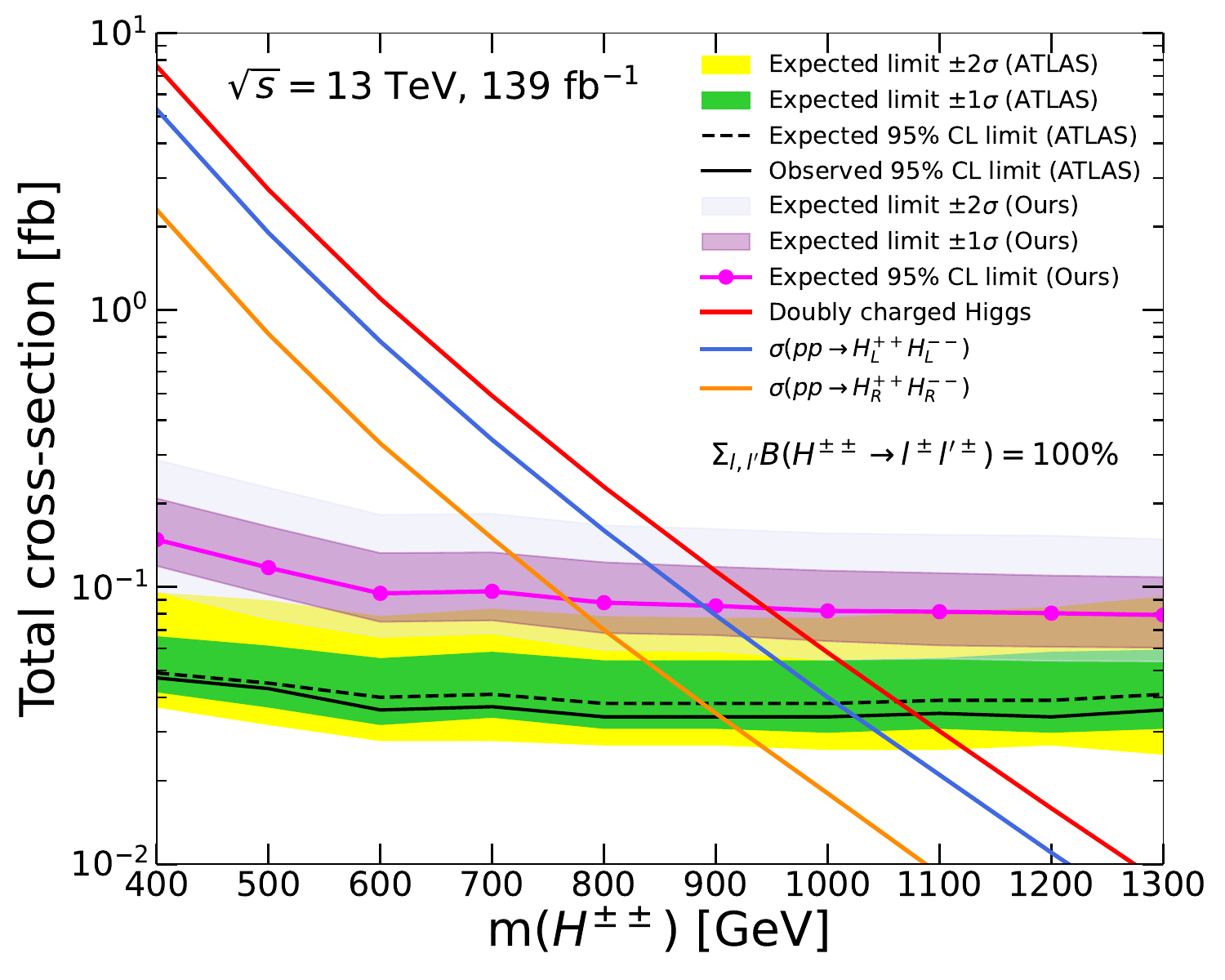}
  \caption{Expected $95\%$~CL upper limit on the
  $pp\to H^{++}H^{--}$ total cross section as a function of
  $\mHpp$ obtained in this work using the corrected four-lepton
  signal yields and the ATLAS background predictions (magenta),
  compared with the ATLAS expected (dashed) and observed (solid)
  limits and their $\pm1\sigma$ (green) and $\pm2\sigma$ (yellow)
  bands. The theoretical Drell--Yan cross sections for $\HppL$
  (blue), $\HppR$ (orange), and their sum (red) are shown with
  their uncertainty bands. All curves assume
  $\sum_{\ell\ell^\prime}\mathcal{B}(\Hpp\to\ell^\pm\ell^{\prime\pm})
  = 100\%$ with equal partial branching ratios.}
  \label{fig:limit}
\end{figure}

Figure~\ref{fig:limit} shows the resulting expected exclusion
limit on the total cross section as a function of $\mHpp$,
overlaid on the ATLAS expected and observed limits and the
theoretical Drell--Yan cross sections for $\HppL$ and $\HppR$.
The expected limit obtained with the corrected signal
yields lies systematically above the ATLAS expected limit across
the entire mass range: it falls from $\sim0.15$~fb at $\mHpp=400$~GeV
to $\sim0.08$~fb at $\mHpp=1300$~GeV, compared with the ATLAS
expected limit of $\sim0.05$~fb to $\sim0.04$~fb over the same
range. As a result, the corrected expected limit intersects the
$\sigma(pp\to H^{\pm\pm} H^{\mp\mp})$ curve at $\mHpp\approx950$~GeV,
roughly $100$~GeV below the ATLAS expected intersection at
$1065$~GeV, and intersects the $\sigma(pp\to\HppR H_R^{\mp\mp})$
curve at $\mHpp\approx770$~GeV, again roughly $100$~GeV below the
ATLAS expected intersection at $880$~GeV.

\section{Conclusions and outlook}
\label{sec:conclusion}

We have reexamined the four-lepton channel of the ATLAS search
for pair-produced doubly charged Higgs bosons in the full Run~2
dataset~\cite{ATLAS:2022pbd}, which provides the strongest limits to
date on the mass of such states. Under the equal-branching-ratio
assumption adopted in that analysis, we derived a strict, mass
independent, theoretical upper bound on the fraction of
pair-production events that can yield four reconstructed light
leptons: at most $0.41$ at truth level, reducing to $0.29$ after
the ATLAS lepton reconstruction efficiencies. The four-lepton
retained fractions implied by the auxiliary cutflow of
Ref.~\cite{ATLAS:2022pbd}, $\gtrsim0.45$ at the tight-lepton stage,
exceed this bound. We showed that the discrepancy cannot be
explained by the misidentification of hadronic $\tau$ decays or
accompanying ISR jets as electrons, which would require fake
rates far larger than realistic values.

Our corrected signal yields give four-lepton acceptances
consistent with the theoretical bound and significantly below
those underlying the ATLAS result. The Ours/ATLAS yield ratio
remains flat at $0.36$--$0.40$ over the full mass range we
tested, consistent with a normalisation offset in the ATLAS
four-lepton samples rather than a mass-dependent physics effect.
This shifts the expected exclusion limit systematically above the
ATLAS expected limit by a factor of at least two across the full
mass range, moving the expected lower mass bound on $H^{\pm\pm}$ from
$1065$~GeV to $\sim950$~GeV within the left-right symmetric
type-II seesaw model, and on $\HppR$ from $880$~GeV to
$\sim770$~GeV within the Zee--Babu model.

Our findings indicate that the four-lepton signal efficiency
adopted in Ref.~\cite{ATLAS:2022pbd} is overestimated. The
corresponding mass limits are widely used to constrain doubly
charged scalars in different BSM scenarios, and may be somewhat
weaker than quoted.

\begin{acknowledgments}
We thank Prof.\ Debajyoti Choudhury for reading the manuscript and
for his useful feedback.
\end{acknowledgments}

\bibliographystyle{apsrev4-2}
\bibliography{references}

\end{document}